Radial dose distribution and effective delta ray radius (Penumbra radius): Determination for some ions passing through water

E. M. Awad [a, *] and M. Abu-Shady [b]

[a,] *Physics Department, Faculty of Science, Menoufia University, Shebin El-Koom, Menoufia 32511 Egypt*

[b,] *Mathematics and Computer Science Department, Faculty of Science, Menoufia University, Shebin El-Koom, Menoufia 32511 Egypt*

[*]*corresponding author e-mail: elsayed_awad@science.menofia.edu.eg (E.M. Awad) , ayawad@yahoo.com, awadem704@gmail.com*

## Abstract

An analytical equation for calculating radial dose of heavy ions in water is introduced by Awad et al. (Applied Radiation and Isotopes, 142 (2018) 135-142). It is simple alternative to Monte Carlo code and is a promising code, however, still needs refinement. Refinement was added through adjusting radial dose integration upper limit which gives the effective $\delta$-ray range, $r_{max}$ and ion's penumbra radius in water as well. Radial dose distributions for 85 ions forming fifteen energy groups from 0.25 to 24 MeV/n were studied. By employing the effective $\delta$-ray range, it was possible to get more consistent radial dose distribution in comparison to experimental and Monte Carlo simulation data. The corresponding *LET* values of those ions were estimated and compared with *SRIM* program. Penumbra radii for 85 ions were determined. Good description for the penumbra radii was obtained using a proposed new equation which fits experimental data as well.

# 1. Introduction

Radial dose distribution around the ion trajectory $D(r,R)$ is the energy deposited per unit mass at distance $r$ normal to the ion pass. It is of principal importance in track structure theory and is significant for many applications in in biology, hadron therapy as well as electronic devices. It can be used for calculating the inactivation cross section of bacteria and enzymes due to heavy ions [2], for estimating ionization quenching in organic plastic scintillator [3], for determine the detection threshold of SSNTDs [4, 5], for expressing the track etchability of some irradiated crystal [6]. Radial dose can also be used to estimate microscopic energy deposition [7], number of clustered DNA damage due to irradiation [8, 9]. Radial dose is utilized for radiation transport software [10, 11], treatment planning for heavy ion radiotherapy, space radiation risk assessment and estimating the cell survival rate for heavy particle cancer therapy [12-14] and predict the cell surviving rate and the relative biological effectiveness (*RBE*) of the ions [15]. Radial dose can be used to determine lateral distribution of charge due to ion imping in silicon memory cells (penumbra radius) which induce soft error and miscalculation in processing unit causing memory cell upset [16].

Many theoretical, semi-empirical analytical models were prosed to explain the interaction of swift heavy ions with matter. The amorphous track model by Katz and co-workers in 60s of the last century [17-19] and its reduction model, the local effect model (*LEM*), by Scholz and Kraft [20, 21] were the founder of such subject. A simple analytical model of ion track structure based on classical collision dynamics were introduced by Kiefer and Straaten [22]. Meanwhile, Chatterjee and Shaefer model [23] is assuming a dense core region of enormous ionization density and a penumbra region where energy deposition occurs mainly in ionization events by energetic secondary electrons released by primary particles. Moribayashi [9] proposing a fitting function that able to calculate the radial dose independent on the incident ion. It is worth noting that radial dose model had many modifications and improvements during the last few decades and still [24-29]. For more information, please refer to these references and references therein.

Monte Carlo (MC) codes for track structure simulations are widely used in radiation and they provide detailed information on microscopic energy deposition in the medium [30-33]. In a calculation of a particle track, the primary particle and its secondary particles are followed in their passage through matter by considering all possible interactions until the particle's energy is totally dissipated. Each consecutive interaction is generated by a random sampling procedure. Most of these codes are dealing with electrons excitation and ionization cross sections. However, another Monte Carlo approach was given in RITRACKS program [34, 35]. The physical and physico-chemical stages of the radiolysis of water due to heavy ions were simulated. The ion is followed on an event-by-event basis, calculating all ionization and excitation events produced and recording the position of the generated radiolytic species and the energy and direction of the secondary electrons. Geant4-DNA [10, 11] is one of the high sophisticated MC codes where the electrons were transported down to 7.4 eV and provide detailed simulation of dose distributions. The radial dose distribution results obtained by the present work will be compared to Geant-4 DNA data.

The aim of the present work is to extend the applicability of Awad et al. [1] code for track structure to calculate the effective $\delta$-ray radius, $r_{max}$. How $r_{max}$ was deduced for every heavy ions in water medium will be explained. The effect of $r_{max}$ on radial dose for 85 ions of different energy groups will be presented. The ions linear energy transfer, *LET* in water will be estimated from radial dose and compared to *SRIM* Ziegler [36] data. The penumbra radii of the different ions that were deduced from effective *δ-ray* radius were compared with the literature data.

## 2. Methodology

A semi-empirical analytical model based on electronic radiation damage is introduced by Awad et al. [1] for calculating the radial dose distribution *D(r,R)*. *D(r,R)* is assumed to be the energy deposited per unit mass in a short cylinder whose axis is parallel to the path of the ion. In this approach, the empirical electron range-energy formula [37] was considered where the electron range, *R* in cm is given as:

$$R(T) = a_1 \left( \frac{1}{a_2} ln \left( 1 + a_2 \frac{T}{mc^2} \right) - \frac{a_3 \frac{T}{mc^2}}{1 + a_4 \left( \frac{T}{mc^2} \right)^{a_5}} \right) \tag{1}$$

$$a_1 = \frac{b_1 A_T}{\rho Z_T^{b_2}}, \qquad a_2 = b_3 Z_T, \qquad a_3 = b_4 - b_5 Z_T,$$

$$a_4 = b_6 - b_7 Z_T, \qquad a_5 = \frac{b_8}{Z_T^{b_9}},$$

where $Z_T$ and $A_T$ are the atomic and mass number of the target, respectively and $b_i$ *(i=1,2,…,9)* are constants independent of absorber material (Table 1).

To calculate the radial dose, two fundamental quantities must be determined. For detail steps and more information, please refer to [1]. The number of liberated electrons in ion-medium interaction and the energy carried by each electron as a function of the radial distance, *r*. Therefore, the electron energy, *T* in MeV as a function of its position must be determined.

To obtain this, the inverse of the energy-range of Eq. (1) must be implemented [37] as follows:

$$T(R) = 0.511 c_1 \left( e^{\frac{R}{c_1} \left( c_2 + \frac{c_3}{1 + c_4 R^{c_5}} \right)} - 1 \right) \qquad (2)$$

Where $c_i$ $(i=1,2...5)$ are constant for a given target $(Z_T, A_T)$ and $d_i$ $(i=1,2,...,9)$ are another constants independent of absorber material and are given in Table 1 as:

$$c_1 = \frac{d_1}{Z_T}, \qquad c_2 = \frac{d_2 Z_T^{d_3}}{A_T}, \qquad c_3 = d_4 - d_5 Z_T,$$

$$c_4 = \frac{d_6}{Z_T^{d_7}}, \qquad c_5 = \frac{d_8}{Z_T^{d_9}}.$$

Equations (1 & 2) are important for defining the necessary constants.

The maximum transferred energy to delta rays is given by:

$$T_{max} = \frac{2mc^2 \beta^2}{1 - \beta^2} \qquad (3)$$

$\beta$ is the ion velocity in medium relative to velocity of light in vacuum. $T_{max}$ can also be estimated [38] as follows :

$$T_{max} = \frac{2\tau(\tau+2)m_e c^2}{1 + 2(\tau+1)\left(\frac{m_e}{M}\right) + \left(\frac{m_e}{M}\right)^2}, \qquad (4)$$

where $\tau$ is the projectile ratio of kinetic energy to rest mass energy and M is the mass of the projectile and $m_e c^2$ is the electron mass energy and E is the kinetic energy of the primary particle. It must be noted that Eq. 3 and Eq. 4 give the same maximum energy transferred to the liberated electrons.

Once the numbers of liberated δ-electrons are found and their energies are assigned, the radial dose can be estimated. It must be remembered that the secondary electrons are assumed to originate from each point on the ion trajectory and the local distributed dose at each radius from the ion trajectory (the radial dose), $D(r, R)$ is therefore as follows:

$$D(r, R) = \frac{-CZ^{*2}}{\rho 2\pi r c_1 \beta^2} \int_{R_{min}}^{R_{max}} \frac{e^{X_4} X_3 (X_2 - X_1) \ e^{\left(\frac{R-r}{c_1} X_1\right)}}{(e^{X_4} - 1)^2} \ dR, \qquad (5)$$

$$X_1 = c_2 + \frac{c_3}{1 + c_4 (R-r)^{c_5}}, \qquad (6)$$

$$X_2 = \frac{c_3 c_4 (R-r)^{c_5}}{(1 + c_4 (R-r)^{c_5})^2}, \qquad (7)$$

$$X_3 = \frac{c_2}{c_1} + \frac{\frac{c_3}{c_4}(1 + c_4 R^{c_5}) - \frac{c_3 c_4 c_5}{c_1} R^{c_5}}{(1 + c_4 R^{c_5})^2}, \qquad (8)$$

$$X_4 = \frac{R}{c_1} \left( c_2 + \frac{c_3}{1 + c_4 R^{c_5}} \right), \qquad (9)$$

where $C$ is a constant and for water (Zhang et al., 1985) equals

$$C = \frac{2\pi N e^4}{mc^2} = 8.5 \ \frac{keV}{mm} = 1.369 \times 10^{-7} \ \frac{erg}{cm} = 1.369 \times 10^{-14} \ \frac{J}{cm}. \qquad (10)$$

For calculating the radial dose, the ion parameters of $Z^*$ (the effective charge number of the ion) which was calculated according to Barkas' formula as $Z^* = Z(1 - e^{-125\beta Z^{-\frac{2}{3}}})$ and $\beta$ were incorporated in the calculation as well as the target parameter of $\rho = 1 \ \frac{gm}{cm^3}$ for water medium.

An algorithm was constructed to calculate the integration in Eq. (5) numerically by using the *Mid-Point Method* of integration. For a given radial distance $r$ slightly less than $R_{min}$ the above integration was estimated between $R_{min}$ and $R_{max}$. Since, the two integration limits $R_{min}$ and $R_{max}$ are significant, $R_{min}$ was kept constant at 0.2 nm. As explained in the previous work [1], $R_{max}$ was deduced from Eq. (3) or (4) and substitute back to Eq. (1). But it was found that calculated radial dose using full $\delta$-ray range does not give accurate dose profile. Therefore, the objective of the current work is to find and propose a newly effective $\delta$-ray range. This effective $\delta$-ray range was determined by the try and error method. This try and error approach was controlled by the next two equations (11 &12) or steps.

The total radial dose distributed over the $2\pi$ around the ion trajectory at all radial distances was determined by carrying out the following double integration.

$$D = 2\pi \int_{R_{min}}^{R_{max}} \int_{r=0.2}^{r=r_{max} \geq R_{min}} r D(r,R) dr dR. \qquad (11)$$

The two integration limits were carried out in such a way that the down limit for the second integration is $R_{min}$ and the upper limit is $R_{max}$ (are the maximum $\delta$-ray range obtained from Eq. 3 or Eq. 4). For the first integration, the lower limit is taken as $r$=0.2 nm and the upper limit of this integration is a part from $R_{max}$ and is taken as $r = r_{max} \geq R_{min}$. The total dose is then determined. This total radial dose is related to the linear energy transfer, $LET$ of the ion at a given energy through the following relation:

$$D = 1.602 \times 10^{-9} \times LET(\frac{keV}{\mu m}) \times fluence(\frac{electron}{cm^2}) \times \frac{1}{\rho(\frac{g}{cm^3})}. \qquad (12)$$

The $LET$ of the ion is then determined. By comparing the obtained $LET$ with the tabulated-$LET$ ($LET_{SRIM}$) of the ions from $SRIM$ program [36], one can judge how accurate the deduced integration limits are. Otherwise, the process carried out for Eq. (11) and Eq. (12) will be repeated many times and every time the suggested integration limit of $r = r_{max}$ will slightly adjusted (increased or decreased) tell *LET* calculated by the present approach comes equal or closer to $LET_{SRIM}$ data. Therefore, the looking for integration limits is determined. This step is used to determine the ion's *LET* as well as the newly suggested $r = r_{max}$ is considered as the ion's penumbra radius.

For radial dose profile determination, the new integration limits of Eq. (5) is then assigned in such a way that $R_{min}$ =0.2 nm and $R_{max}$ will become the newly suggested effective *δ-ray* range   $r = r_{max}$. Then *r* increased step-by-step smoothly with 1 nm long and the corresponding $R_{min}$ was adjusted accordingly. The above radial dose integration was carried out at every radial distance, *r* i.e. at *r*=1 nm then at r=2 nm and so on with 1 nm step till the end of the effective *δ-ray* range reached (an extensive work). The integration limits were determined for the 85 ions under study before calculating the radial dose distribution for any ion. A *FORTRAN* algorithm was constructed to facilitate the above integrations numerically by using the *Mid-Point Method* of integration.

### 3. Penumbra radius

According to the track structure theory, the energy deposited by the ion is consists of dense core surrounding by a region called penumbra. Penumbra is the region where secondary electrons released by the primary particle deposit their energy. It shows how far the secondary electrons reach its effect around the ion trajectory and the damage it may cause for biological as well as electronic systems. The outer border (radius) of penumbra [23] is given by

$$r_{max} = 0.768E - 1.925\sqrt{E} + 1.257 \quad [\mu m] \qquad (13)$$

Another penumbra radius formula [22] is given as

$$r_{max} = 6.16 \times 10^{-2}E^{1.7} \quad [\mu m] \qquad (14)$$

Where, $E$ is the kinetic energy of the projectile [MeV/nucleon]. The penumbra radii obtained by the present work ($r = r_{max}$) are compared with the data obtained by those equations.

## 4. Experimental and Geant-4 DNA data

Radial dose distribution was measured experimentally by tissue-equivalent gas chamber. A large ionization chamber consists of an aluminum cylinder and a copper central wire serves as an ion collector. A probe is attached to rotate into various radial positions from the center to almost the chamber wall. A faraday cup was used to determine the beam intensity and ionization current was measured using a vibrating read electrometer where ionization was converted to energy deposited. The experimental data [39-42] for 1 MeV proton, 3 MeV alpha, 24 MeV Carbon and 41.1 MeV Oxygen were collected and used for comparison with present calculations.

Geant4-DNA toolkit data for p, α-particles, C and O ions of different energies were obtained [11] and compared with current calculations. In Monte Carlo simulation, these ions were shot into liquid water and energy depositions around the particle track were scored in concentric cylindrical shells around the incident particle track. Each shell has a thickness of 1 nm. The validation of the current approach was achieved through comparing the present calculations with the available experimental as well as Geant-4 *DNA* data. Table 2 compile the fifteen groups of 85 ions of different E/n where the experimental data are defined as ($^{*}$) and Monte Carlo Data are defined as ($^{**}$) and there references.

## 5. Results

### 5.1 δ-ray effective range, $r_{max}$

Integration limits play an important role in radial dose estimation especially the upper limit as explained in the Methodology section. The effective range for 85 ions representing 15 groups, see Table 2 were determined. The ratio between the effective δ-ray range, $r_{max}$ (Eqs. 11&12) and maximum δ-ray range of the liberated δ-electrons, $R_{max}$ (Eqs. 3&1) ($R_{ratio} = \frac{r_{max}}{R_{max}}$) were determined for the different energy groups and are illustrated as a function of the group velocity, $β$ in Fig. 1. One can observe that there are three different $R_{ratio}$ regions, one before 1 MeV/n ($β$=0.046), one between 1 to 15 MeV/n and one from 15 MeV/n and above. At energy equal or less than 1 MeV/n, the effective radii for those ions are about constant and their average is $\approx \frac{1}{3} R_{max}$. For ions of energy greater than 1 MeV, the effective radii increase by increasing the ion energy. At energy equal and greater than 15 MeV/n, $R_{ratio}$ becomes closer to 1 and the effective radius, $r_{max}$ will almost equal full δ-ray range, $R_{max}$.

The validity of the current calculations as well as accurate integration limits determination are guaranteed by the fact that the radial dose integration over all radial distances around the ion path must yield the ion's *LET* (Eqs. 11 &12). Fig. (2) shows the ratio between the calculated linear energy transfers to the tabulated ones, $LET_{ratio} = \frac{LET}{LET_{SRIM}}$ for the fifteen groups under study. Please note that the vertical dashed line at 1 MeV/n and the horizontal dashed line at 1 of y-axis are to guide the eye only. One can observe that, there are two different regions for the calculated linear energy transfer. One bellows 1 MeV/n in which the calculated *LET* is less than $LET_{SRIM}$ by 30% and one above it in which the two *LETs* are about equal. It seems that the current model underestimates the impeded energy, the impeded radial dose and hence the corresponding ion's *LET* for low energy ions (<1MeV/n). However, the present approach gives an excellent calculations for the impeded energy, the impeded dose as well as the ion *LET* in the region above 1 MeV/n where the $LET_{ratio} = \frac{LET}{LET_{SRIM}}$ exactly equal 1. One has to say that, this $LET_{ratio}$ study is the most significant step in such studies.

## 5.2    *Radial dose*

Once the desired effective *δ-ray* range, $r_{max}$ is determined, the integration limits of Eq. 5 are assigned and the radial dose can be calculated at every position, *r* from the ion trajectory. Implementing the new integration limits on radial dose estimations has improved the dose profile. Fig. 3 shows the improvement in radial dose profile of the some ions by using current approach (using $r_{max}$) in comparison with profiles of using the full *δ-ray* range as well as Geant-4 DNA Monte Carlo data. To distinguish between data, data for 1 MeV P (Full, effective and Geant-4) are divided by 100, data for 0.75 MeV/n alpha are divided by 10, data for 2 MeV/n C are kept the same while data for 2.569 O are multiplied by 10. Good improvements were obtained especially for proton and alpha where exact matching between current radial dose and Geant-4 DNA radial dose data at almost all radial distances was observed. Reasonable improvements were found for C and O profile and C data is better because it becomes closer to Geant-4 data but no exact matching at larger radii were observed.

Radial dose profile using $r_{max}$ for a big data set of 85 ions divided into 15 groups were studied. The radial dose profile for some of these groups will be shown and the other was not for similarity only. The group that has experimental and/or Geant-4 *DNA* data will be presented in this context. The calculated radial dose profiles for P, D, α, C, O, I ions at energy equal 0.5, 1 and 2 MeV/n as a function of the radial distance, *r* are presented in Fig. 4 (a-c). The experimental data for D as well as I are dropped in Fig. 4-a. Experimental data for P, D and P Geant-4 data are given in Fig. 4-b. While experimental data for P, C and C Geant-4 are given in Fig. 4-c for the sake of comparison. Despite the current approach underestimates the impeded dose of ions at energy 1 MeV/n and below as stated before, excellent dose profile and good matching between the current approach, experimental as well as Monte Carlo data. Similarly, the dose profile for P, α, C, O, Ne and Fe as a function of radial distance, *r* at energies 2.4, 3 and 8.1 MeV/n are compiled in Fig. 5(a-c). Experimental data for O, P and Ne are added to the figures.

$D(r,R) \times r^2$ as a function of radial distance, *r* discerns the differences between the different dose profile obtained by current approach, experimental and Geant-4 data. Fig. 6 (a-c) shows that calculated dose profile, experimental and Geant-4 data are following the $\frac{1}{r^2}$ behavior in agreement with previous studies [12, 22, 43-45]. Therefore, the spatial distribution of the energy deposited is not homogeneous, but each individual ion deposits its energy with an approximately $\frac{1}{r^2}$ dependence on the radial distance from the trajectory.

Radial dose profile for ions P, α, C, O and Fe at energy 10 MeV/n in comparison with Geant-4 DNA MC simulation data are presented in Fig. 7-a. The two dotted vertical lines at $r=1$, 4 nm are to guide the eyes only. To improve the comparison between the two model, the values of $D(r,R) \times r^2$ for these ions as a function of the radial distance, $r$ are given in Fig. 7-b. Monte Carlo data show better $\frac{1}{r^2}$ behavior than the present approach at such high energy. One can observe that in the region of r < 1 nm, the radial dose is slightly underestimated while in the region between $r = 1$-4 nm the present calculations and Monte Carlo simulation are almost similar and after 5 nm radial distance the deviation between the two data increases. At radial distance equal 100 nm the current radial dose gives larger dose than Geant-4 data.

## 5.3    Radial dose applications

The energy deposition radially around the ion trajectory and the corresponding distribution of dose around the ions can be considered as a fundamental quantity and can be used to deduce some important information about the incident ion in the medium. From radial dose one can determine the stopping power and/or the linear energy transfer (*LET*) of the incident ion as well as how far from the ion is the energy deposited effectively (the ion's penumbra radius).

### 5.3.1  Ion LET

Estimating the total radial dose distributed over the $2\pi$ around the ion trajectory at all radial distances was determined by the double integration of Eq. 11. The corresponding ion LET is thus directly deduced from Eq. 12. The LET (KeV/µm) of the six ions under investigation (P, α, O, C, Ne and Fe) at the energy range from 0.25 to 24 MeV/n in comparison with *SRIM* as well as Benton-Henke range-energy Table [46] are shown in Fig. 8 (a-c) using logarithmic scale. For better illustration, two of these ions (O and Fe) are showing as example in Fig. 9 (a, b) using linear scale. Investigating these figures one can observe that *LET* as a function of the ion's energy can be split into two separate parts; before the Bragg peak (Bethe–Bloch) and after the Bragg peak (Thin Down) [47]. In the Bethe-Bloch region, the model is able to determine the ion's *LET* exactly like *SRIM* and Benton-Henke Table but in the Thin Down region the accuracy of determine the ion's *LET* is not the same.

### 5.3.2 *Penumbra radius*

After adjusting the integration limits as was explained, the upper integration limit or the effective *δ-ray* range, $r_{max}$ can be determined. This $r_{max}$ is considered as the ion's penumbra radius. The average penumbra radius at the studied energy from 0.25 to 24 MeV/n was calculated and presented in Fig. 10. The statistical error in each point is smaller than the circle symbol given in the figure. It was found that ion's penumbra radius; $r_{max}$ depends on the ion's kinetic energy, *E* (energy/n) through the following polynomial equation

$$r_{max} = 0.068E^2 + 0.2215E - o.2339 \ \ (\mu m),$$
$$R^2 = 0.9999 \tag{15}$$

Comparing the obtained penumbra radius with penumbra radii obtained by Kiefer and Straaten [22] (Eq. 13) and by Chatterjee and Schaefer (Eq. 14) [23]one found that they all are in the same order of magnitude and the current approach gives larger penumbra radius than the other equations do. The present approach predicts the penumbra radius may be more accurately than the other models. The lateral distribution of the charge generated in silicon by 15 MeV oxygen ion was determined experimentally [16] and it was found equal 300-400 nm in agreement with the present model as shown in Fig. 10.

## 6. Discussion

Modified radial dose profile by taking into account the effective $\delta$-ray range has improved the obtained radial dose for 85 ions of different energies. The results of the present approach are compared against experimental and Geant-4 DNA data. Global agreement was found between the different data at all radial distances as illustrated in Figs (4-7). At the ion's core and at smaller radii of less than or equal 5 nm good matching and coincident between the present calculations, experimental and Geant-4 DNA data was obtained. In the core region a huge number of ionized electrons interacting with each other and with the surrounding electrons where most of the ionizations occur in a small cylinder close to the ion path. In addition to this, slow secondary electrons may be trapped by track potential which formed by the electric field near the incident ion path [48, 49] creating high doses. The present approach succeeded to predict such high dose in this region.

But at high energy ions >8 MeV/n and at higher radial distance > 50 nm, the present approach overestimates the radial dose and the $\frac{1}{r^2}$ behavior is not exactly fulfill. The interaction of high-energy ions is characterized by almost pure electronic excitation and ionization of the target atoms. The primary ionization and excitation processes and the following electron cascade have stochastic nature. Therefore, the assumption that all $\delta$-electrons are liberated normal (with $\approx 90^{o}$) to the ion path may not an appropriate assumption and needs to be revised and angular percentage of $\delta$-electrons must be considered. At large distances, the dose is mainly due to the slowing down of energetic electrons and the model is not able to recognize such slowing rate. The reason for this is not clear but may be the $\delta$-ray range, number of collisions as well as the rate of electron energy loss are not easy to be exactly determined.

Effective $\delta$-ray range, $r_{max}$ increases by increasing the ion's energy (Fig. 1) and $r_{max}$ is always less than $R_{max}$. The reduction in $\delta$-ray range could be  accepted due to the stochastic nature of $\delta$-electron motion in the surrounding medium [35]. The predicted radial dose distribution has improved by using the new proposed $\delta$-ray range, $r_{max}$. However, ions of $E \leq 1$ MeV/n are not following the general trend where $r_{max} \approx \frac{1}{3} R_{max}$. The fewer energy ions are producing less $\delta$-ray energy. These liberated electrons of less energy may suffer larger scattering and resistance with the surrounding electron matrix producing less effective $\delta$-ray range. This may explain the smaller effective $\delta$-ray range in this region.

The deduced linear energy transfer *LET* of the studied ions (85 ones) by the current approach are excellent in the Bethe-Bloch region of Bragg peak where the obtained *LET* of the ions are almost equal *SRIM* and Benton-Henke data, see Figures 8&9. But in the Thin Down region, the model underestimate the ion's *LET* by 30% as stated before.  In Thin Down region of Bragg's peak the ion's characteristic (effective charge, range, straggling, *LET* etc.) and behavior is always difficult to be predicted. May be excitation electron energy and the elastic collision between the incident ions with surrounding medium must be taken into account. Bellow 1 MeV/n elastic collision may dominant causing underestimation for the impeded energy as well as the impeded dose. Energy of the produced free radicals energy must be considered as well [34].

Adjusting the integration limits lead to the deduced Penumbra radius of the studied ions, Fig. 10. As expected, as the ion's energy increases as the liberated $\delta$-electron energies increases (Eq. 3 & 4) leading to an increased penumbra radius. Penumbra radius obtained by the present work and other researcher having the same order of magnitude, however, it is 3 times larger the penumbra radii deduced by other formula. Fortunately, the experimental data for the lateral distribution of charge due to oxygen in silicon [16] is closer to the suggested formula in this work. Experimental data for lateral distribution of charge for the different ions of different energy are needed for further examination of the present formula.

## 7. Conclusion

Radial dose distributions of different ions of different energies in water have been improved by applying the effective $\delta$-ray range, $r_{max}$. Effective $\delta$-ray range was deduced by adjusting the radial dose integration limits. Detail steps were given and the final radial dose integration was solved numerically using the Mid-Point Method. Radial dose for P, D, alpha, C, O, I, Ne and Fe at energies 0.25, 0.50, 0.75, 1, 2, 2.4, 2.569, 3, 4, 6, 8.1, 10 and 24 MeV/n were determined. The model was able to reproduce successfully the radial dose for such ions compared to experimental as well as the Geant-4 DNA. The variation in radial dose multiplied by $r^2$ $(D(r) \times r^2)$ as a function of the radial distance, *r* was estimated and the variation between the different calculations are clearly demonstrated. All ions are following the inverse square relation $\frac{1}{r^2}$ where all ions showing straight line in agreement with experimental as well as Monte Carlo estimation.

The model is able to predict the ion's *LET* exactly compared to *SRIM* and Benton-Henke data in the *Bethe-Bloch* region of Bragg's peak. However, in the *Thin down* region it found some difficulty to determine it exactly.

Penumbra radii in water for the 85 ions understudy were determined. New formula for penumbra radius is suggested in the present work and tested experimentally and compared to other formula. However, the newly suggested formula needs further examination by new experimental data.

The present model is offering a simple alternative to time consuming Monte Carlo simulations and can be conveniently used in hadron therapy dosimetry. It can be used for estimating the radial dose of the ion; it's *LET* as well as lateral distribution of charge around the ion's path (its penumbra radius).

Despite the success obtained by adding the effective *δ-ray* range, the present approach still suffering some limitation especially at the larger radii where the $\frac{1}{r^2}$ behavior is not exactly fulfilled. Radial dose distribution is of particular significance for improving the understanding of radiobiology experiments with heavy ions, radiation protection issues and radiation risks associated for heavy ion exposure. Therefore, further refinement still required where angular distribution of δ-ray must be considered and incorporated into calculations.

Latent track structure in PADC and other solid state nuclear track detectors will be examined using the current approach. The assumption that ions of the same *LET* can have completely different radial dose and different penumbra radii must be tested in the future using current approach.

**Acknowledgment**

**Figure Captions**

Fig.1    The variation in range ratio, $R_{ratio} = \frac{r_{max}}{R_{max}}$ as a function of the ion's relative velocity, $\beta$ for the different energy groups.

Fig.2    The relative $LET$ ratio, $LET_{ratio} = \frac{LET}{LET_{SRIM}}$ of the different ions and groups as a function of the ion's kinetic energy, $E$ (MeV/n).

Fig.3    Calculated radial dose for 1 MeV P divided by 100, 0.75 MeV/n α divided by 10, 2 MeV/n C and 2.569 MeV/n O multiplied by 10 using the full range, $R_{max}$ and the effective δ-ray range, $r_{max}$ compared to Geant-4 DNA simulation data at the different radial distances, $r$.

Fig.4(a-c)    a- The calculated radial dose distributions for P, D, α, C, O, I and I(EX) for the group energy 0.5 MeV/n. b- The calculated radial dose distributions for P, P(EX), P(Geant-4), D, D(EX), α, C, O, Ne and Fe for the group energy 1 MeV/n. c- The calculated radial dose distributions for P, P(EX), α, C, C(EX), C(Geant-4), O, Ne and Fe for the group energy 2 MeV/n, respectively.

Fig.5(a-c)    a- The calculated radial dose distributions for P, α, C, O, O(EX), Ne and Fe for the group energy 2.4 MeV/n. b- The calculated radial dose distributions for P, P(EX), α, C, O, Ne and Fe for the group energy 3 MeV/n. c- The calculated radial

dose distributions for P, α, C, O, Ne, Ne(EX) and Fe for the group energy 8.1 MeV/n, respectively.

Fig.6(a-c)  a- The $Dose{\times}r^2$ for P, P(EX), P(Geant-4), D, D(EX), α, C, O, Ne and Fe as a function of the radial distance, $r$ for the group energy 1 MeV/n. b- The $Dose{\times}r^2$ for P, P(EX), α, C, C(EX), C(Geant-4), O, Ne and Fe as a function of the radial distance, $r$ for the group energy 2 MeV/n. c- The $Dose{\times}r^2$ for P, α, C, O, Ne, Ne(EX) and Fe as a function of the radial distance, $r$ for the group energy 8.1 MeV/n, respectively.

Fig.7(a,b)  a- Radial dose distribution P, α, C, O, and Fe compared to the corresponding Geant-4 DNA data. b- The $Dose{\times}r^2$ for P, α, C, O, and Fe compared to the corresponding Geant-4 DNA data for the group energy 10 MeV/n, respectively.

Fig.8(a,b)  $LET$ calculated by the present approach for (P, α and O) and (C, Ne and Fe), respectively at the different ion's energy compared with the corresponding ones obtained by Benton-Henke energy table and $SRIM$ code.

Fig.9(a,b)  Bragg's peak for O and Fe ions where $LET$ as a function of Energy (MeV/n) are given, respectively. The model exactly fits the data in the Bethe-Bloch region however the accuracy in the Thin Down region is not the same.

Fig.10    Illustrates the predicted penumbra radius as a function of the ion's energy for 85 ions under investigation and the deduced formula in comparison with Kiefer-Straaten and Chatterjee-Shaefer formula.

**Table Captions**

Table 1

Values of the constants $b_i$ and $d_i$ used for electron range-energy equation, Eq. (1) and its inverse relation of energy-range equation, Eq. (2).

Table 2

Radial dose distribution for fifteen groups of 85 ions of different E/n were studied, some of them has experimental data ($^*$) and others has Geant-4 Monte Carlo Data ($^{**}$) for the sake of comparison.

Fig. 1

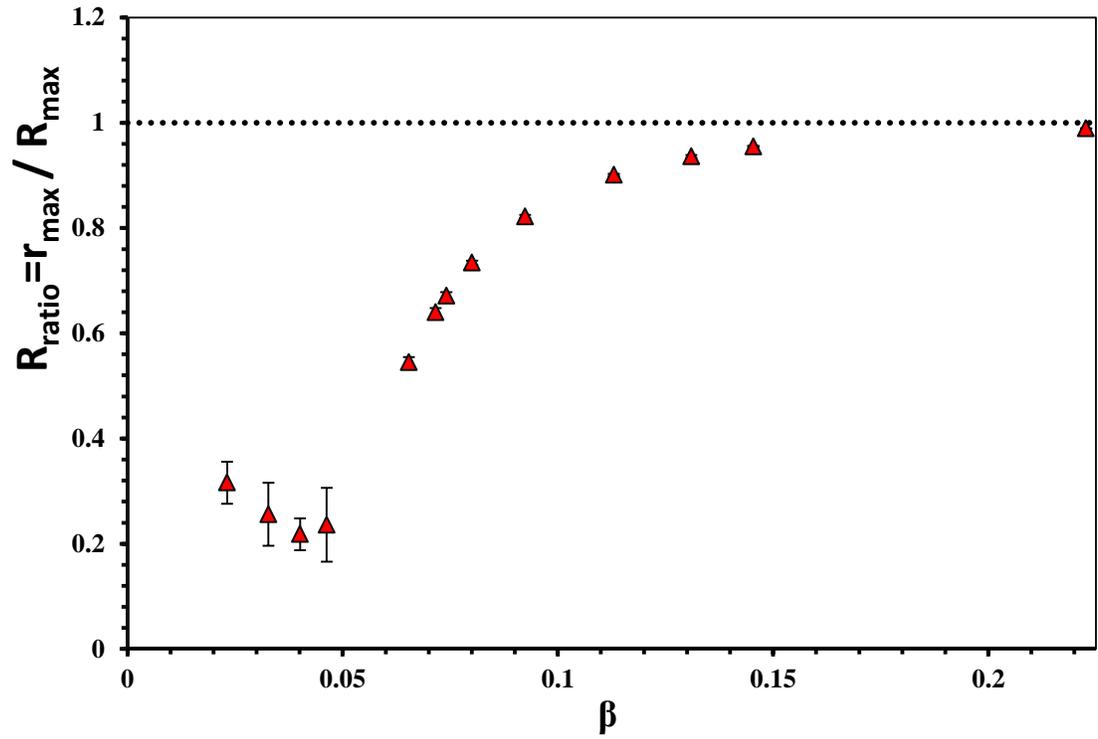

Fig. 2

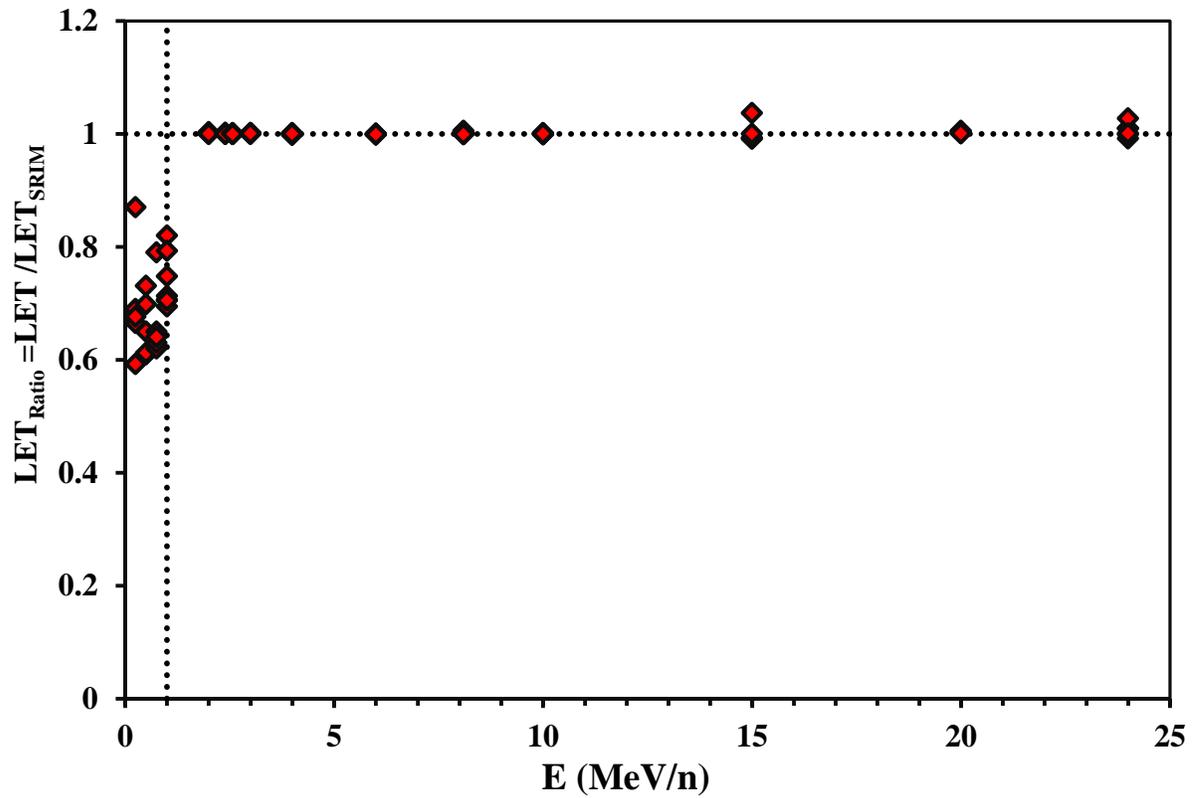

Fig. 3

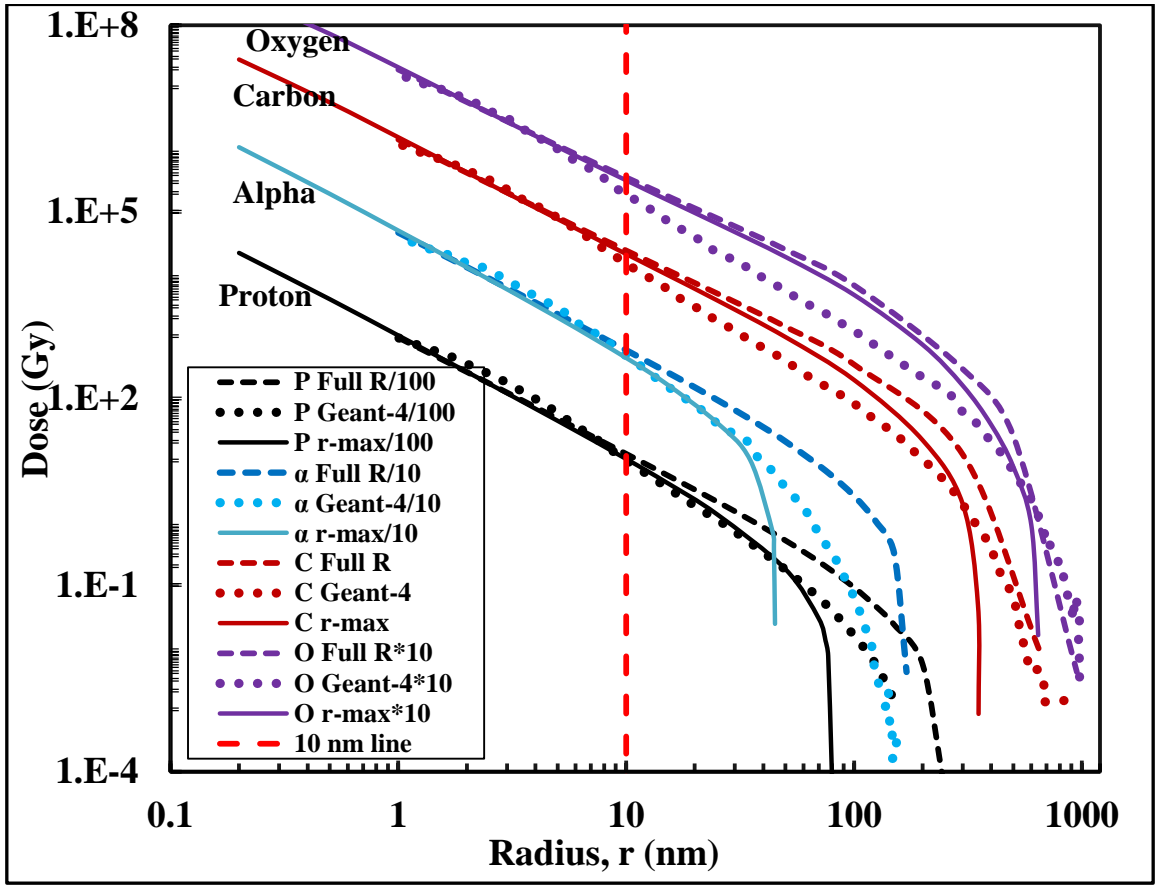

Fig. 4-a

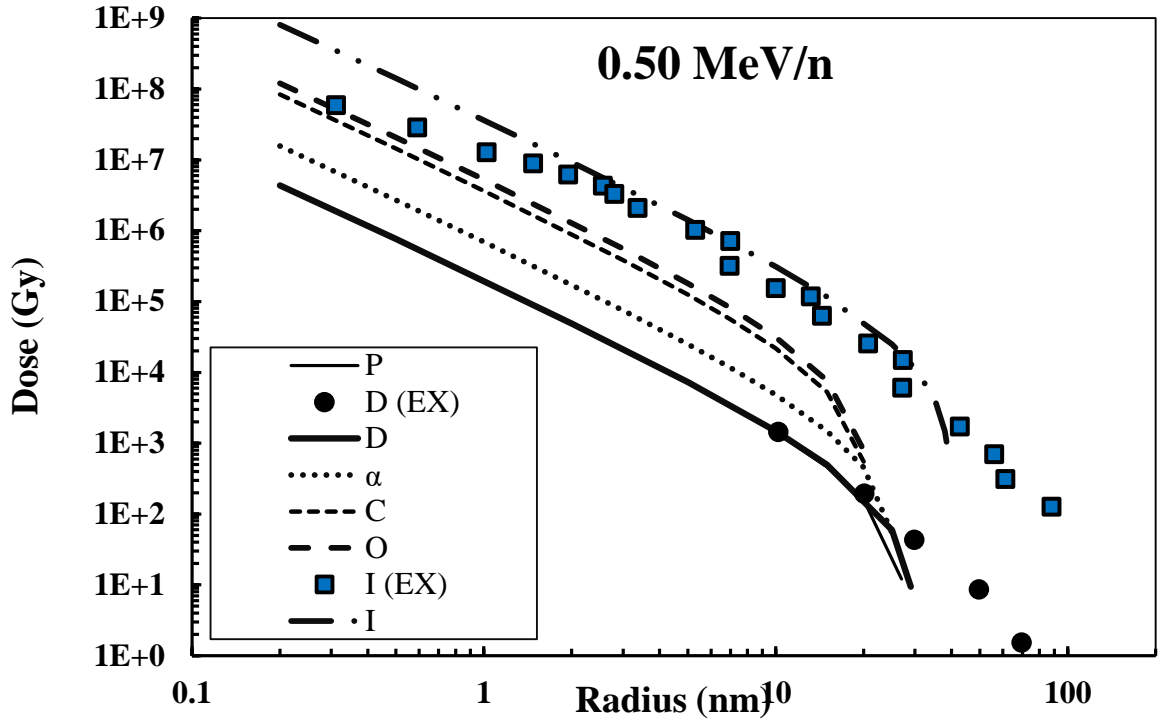

Fig. 4-b

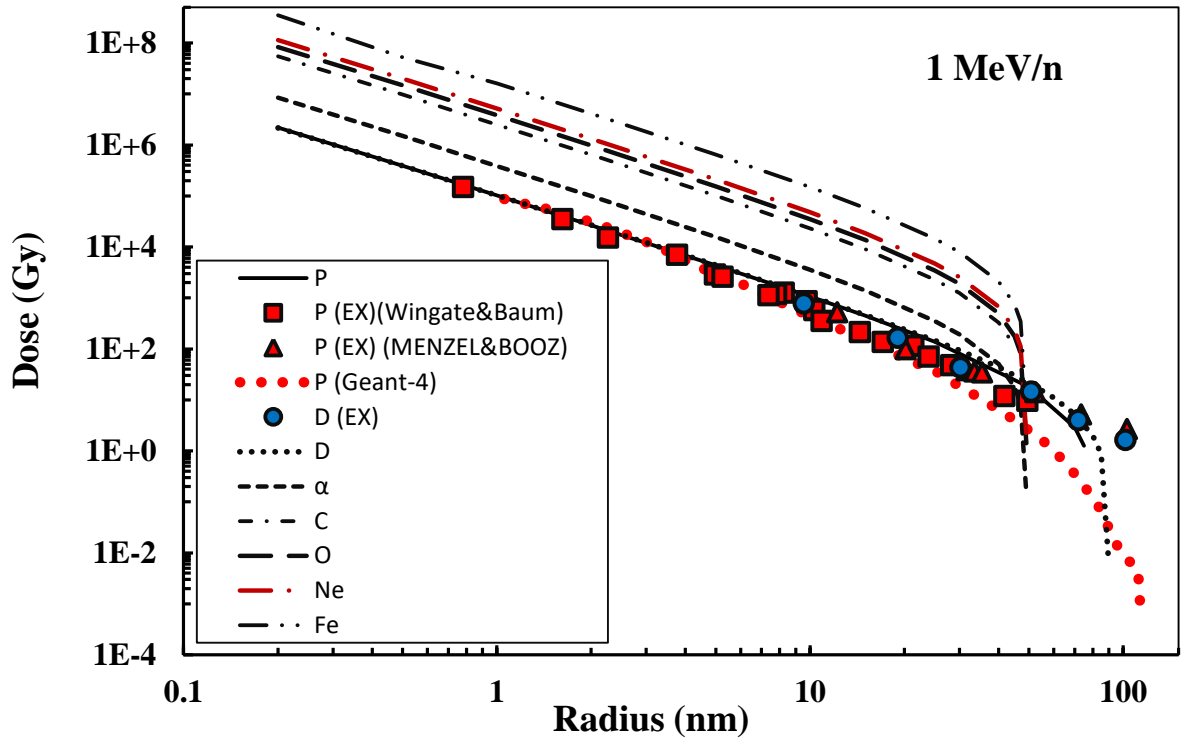

Fig. 4-c

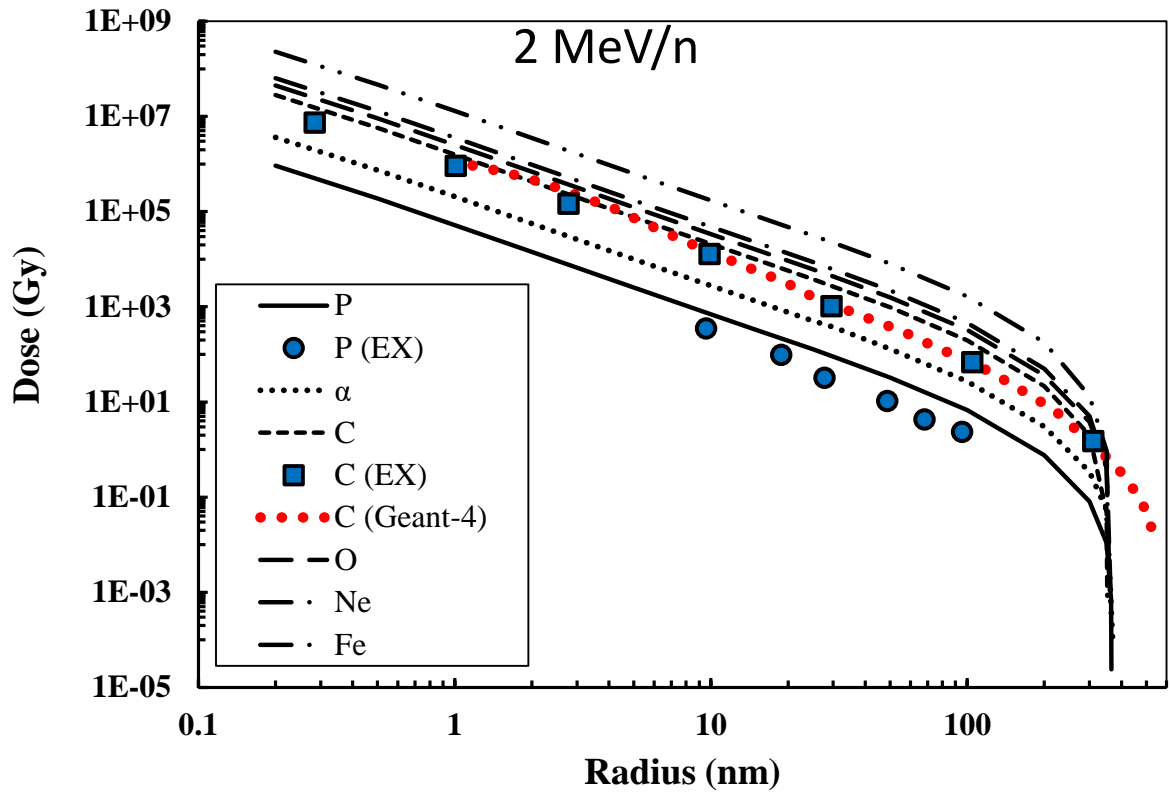

Fig. 5-a

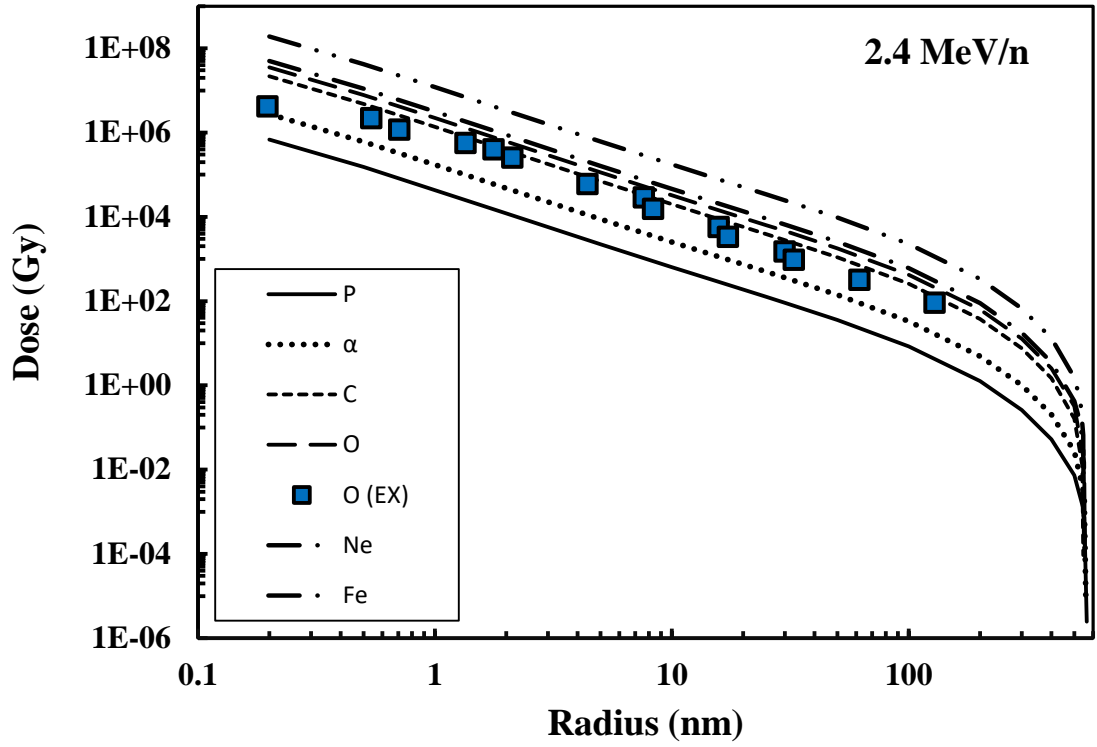

Fig. 5-b

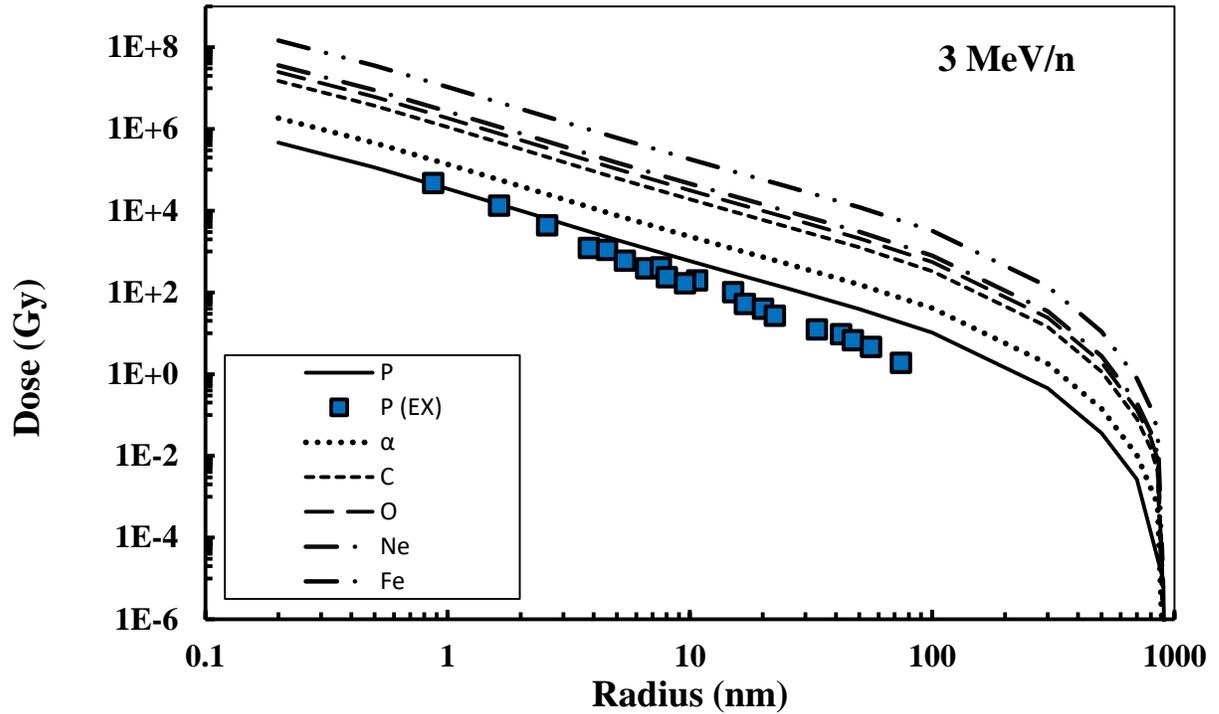

Fig. 5-c

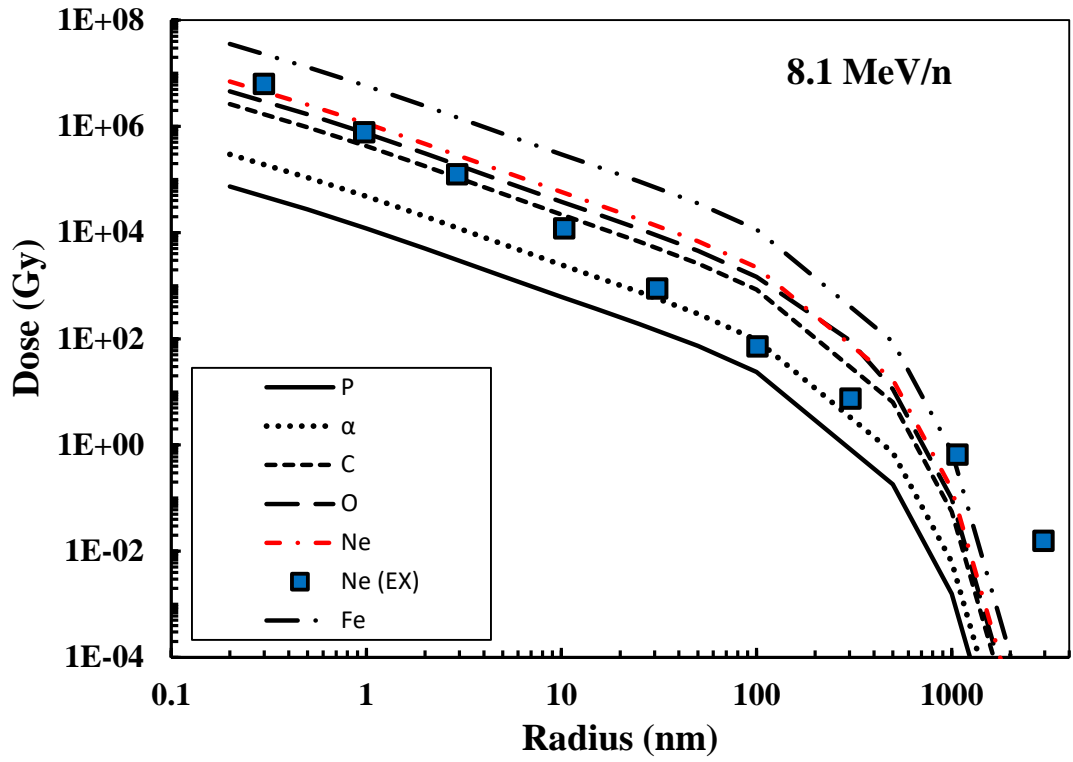

Fig. 6-a

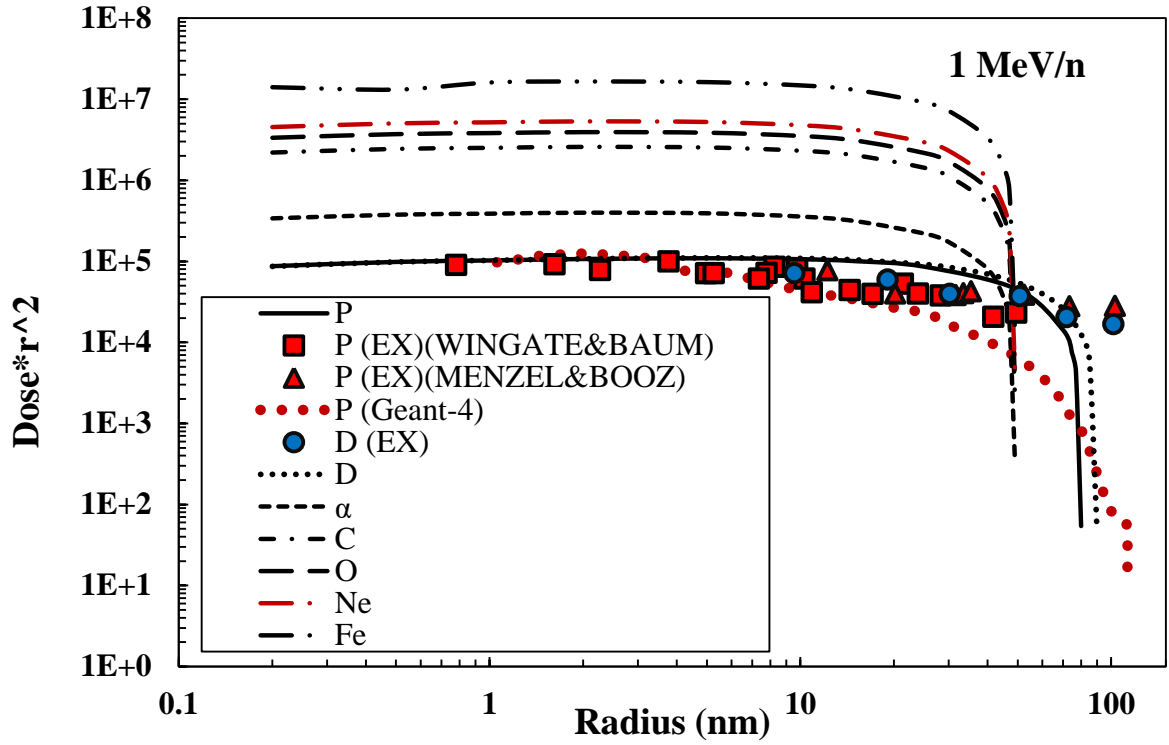

Fig. 6-b

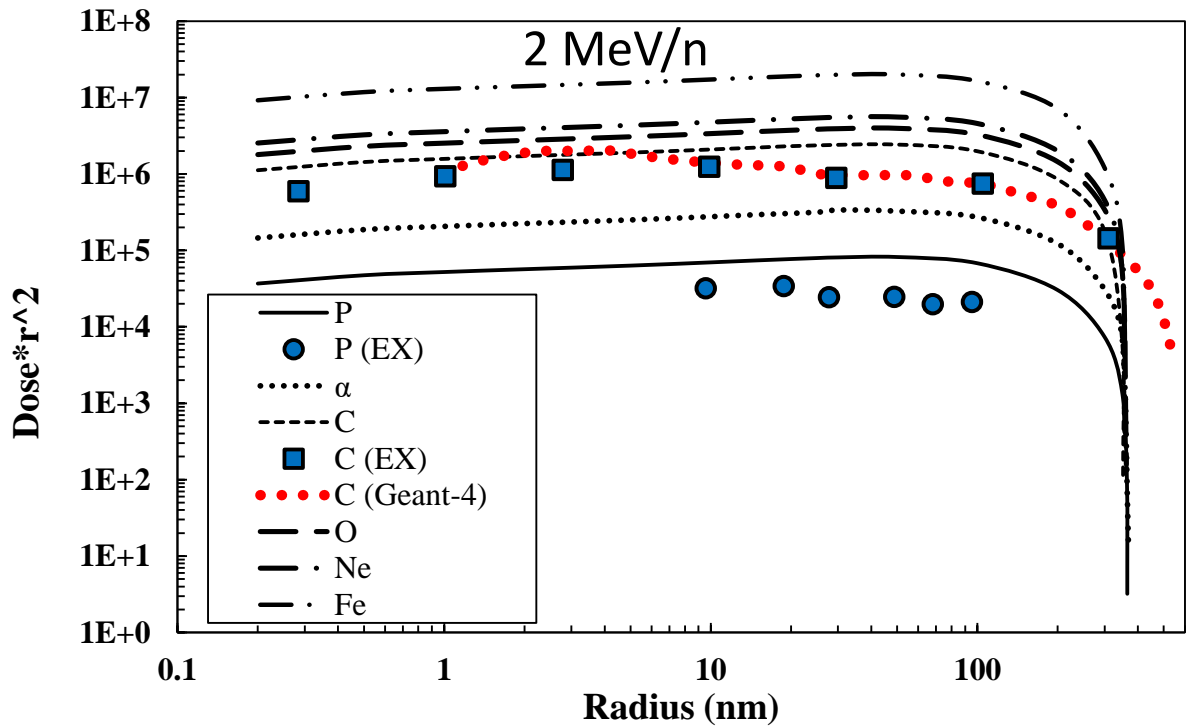

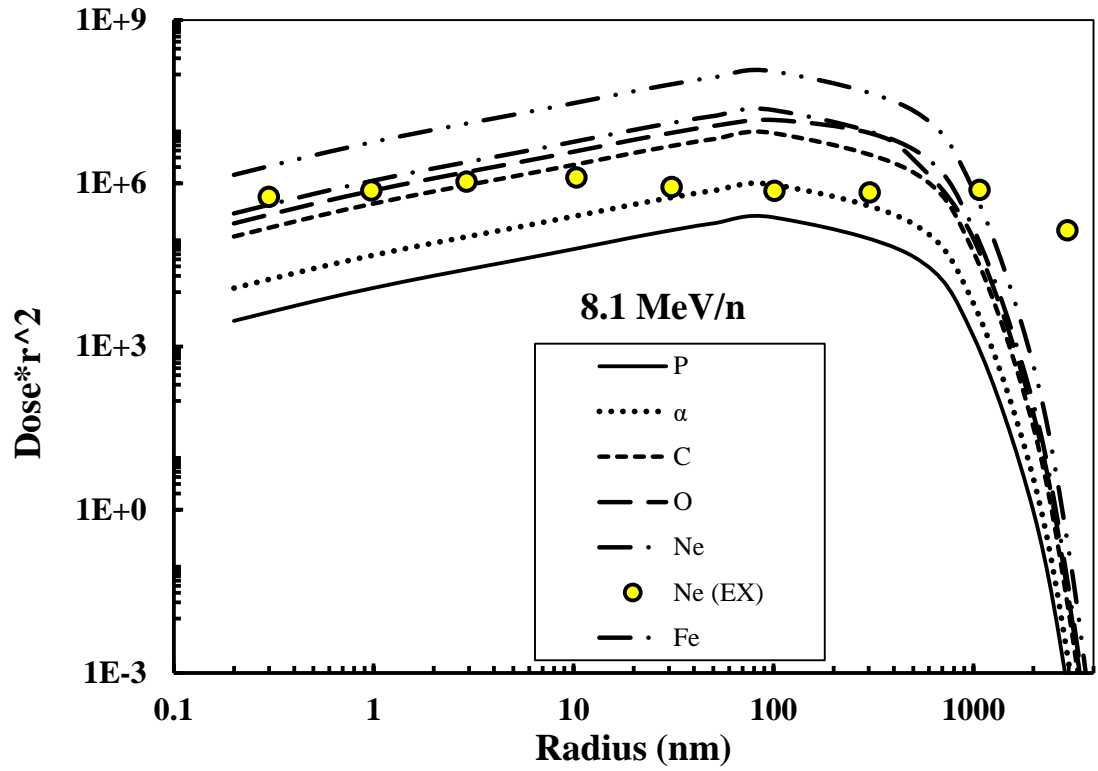

Fig. 7-a

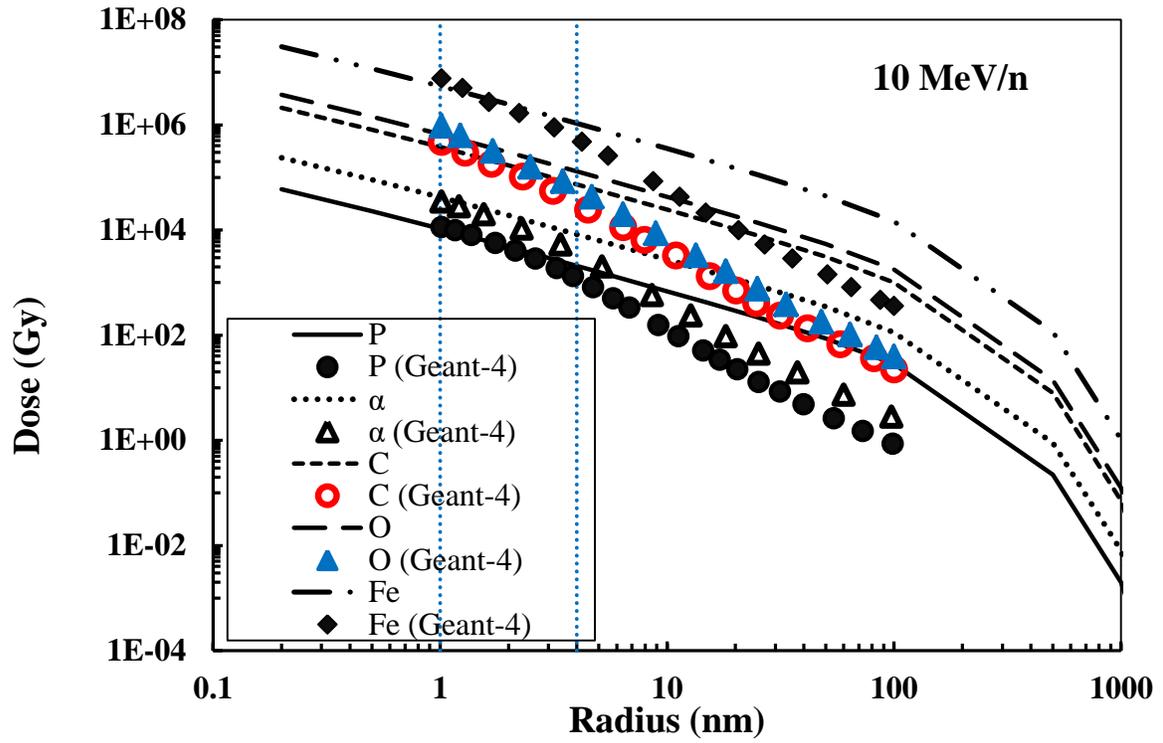

Fig. 7-b

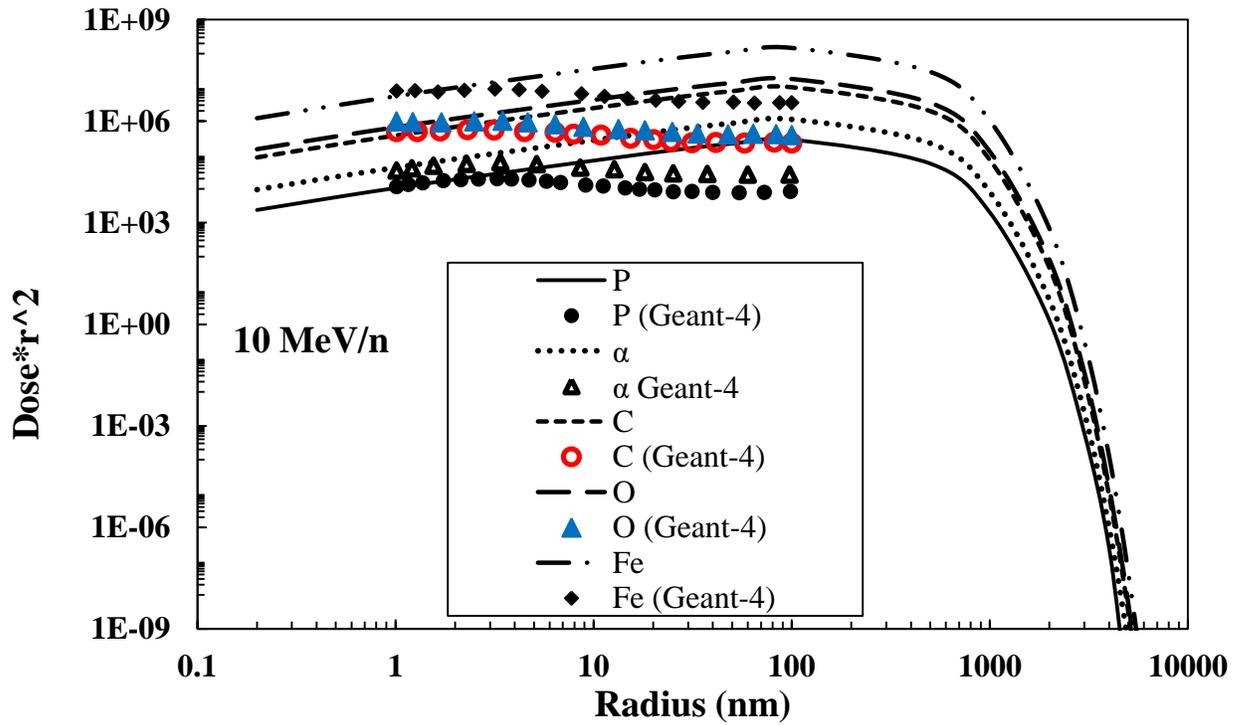

Fig. 8-a

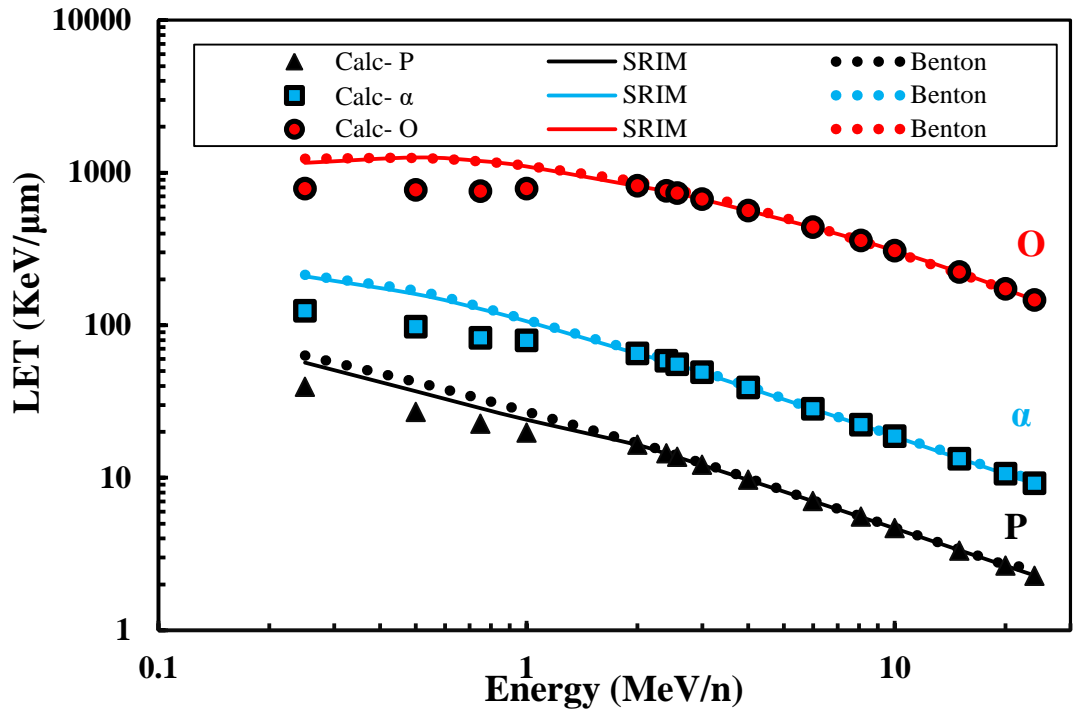

Fig. 8-b

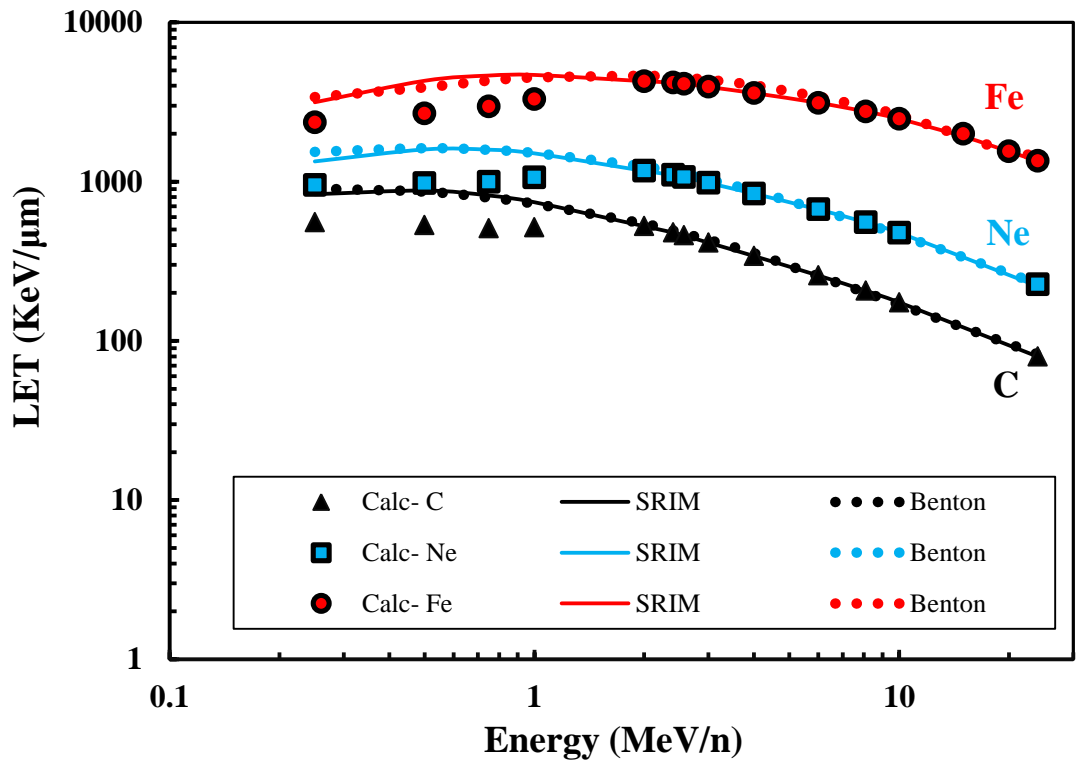

Fig. 9-a

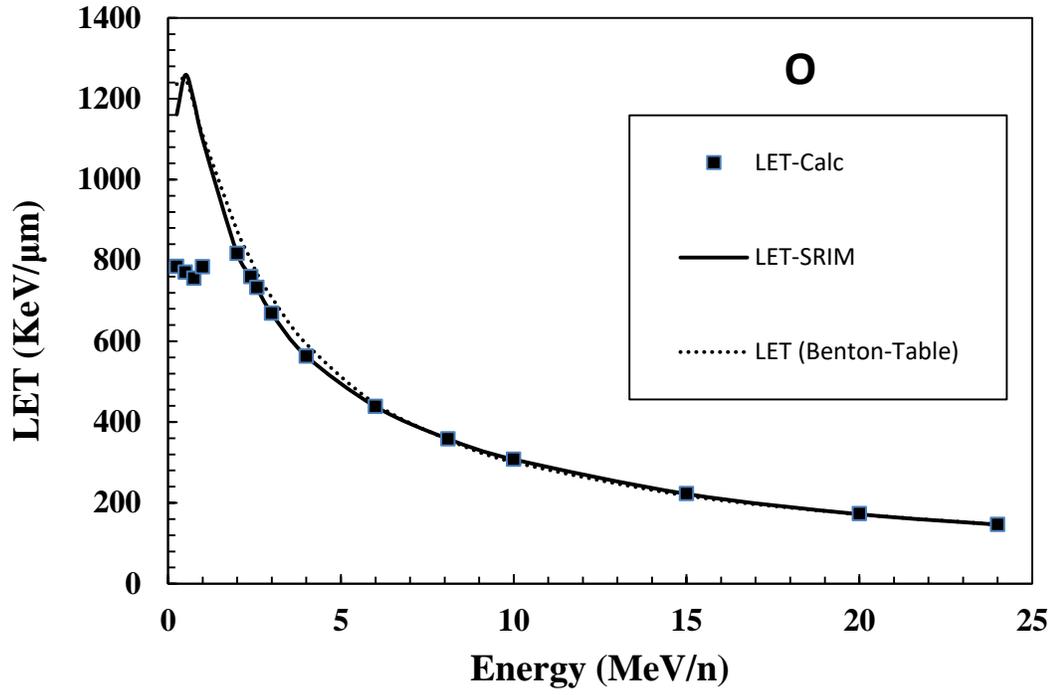

Fig. 9-b

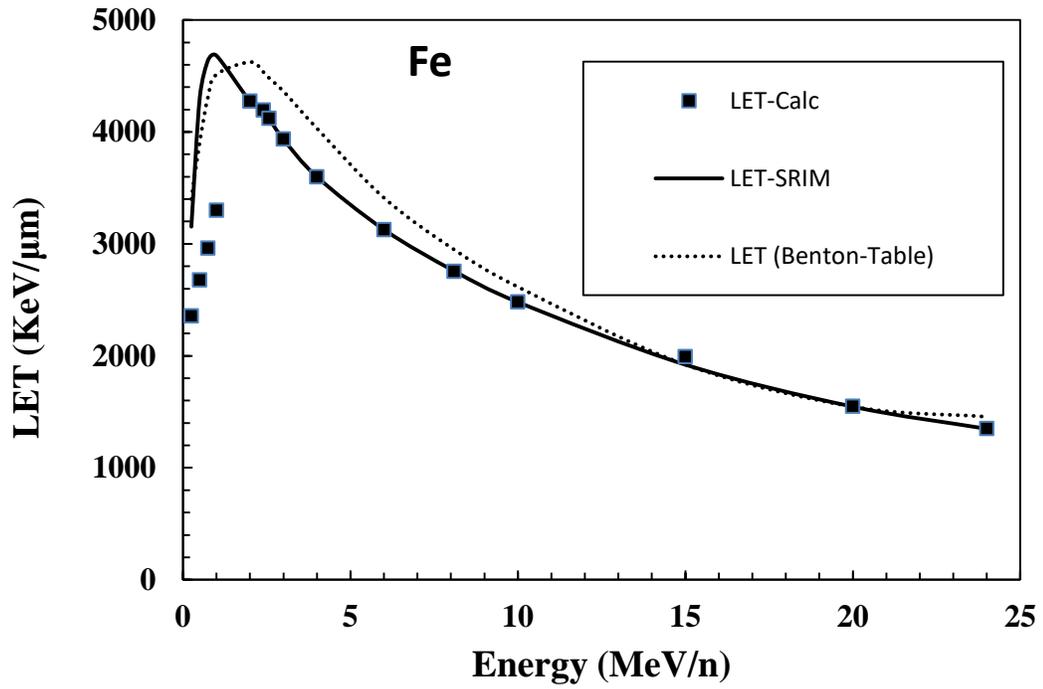

Fig. 10

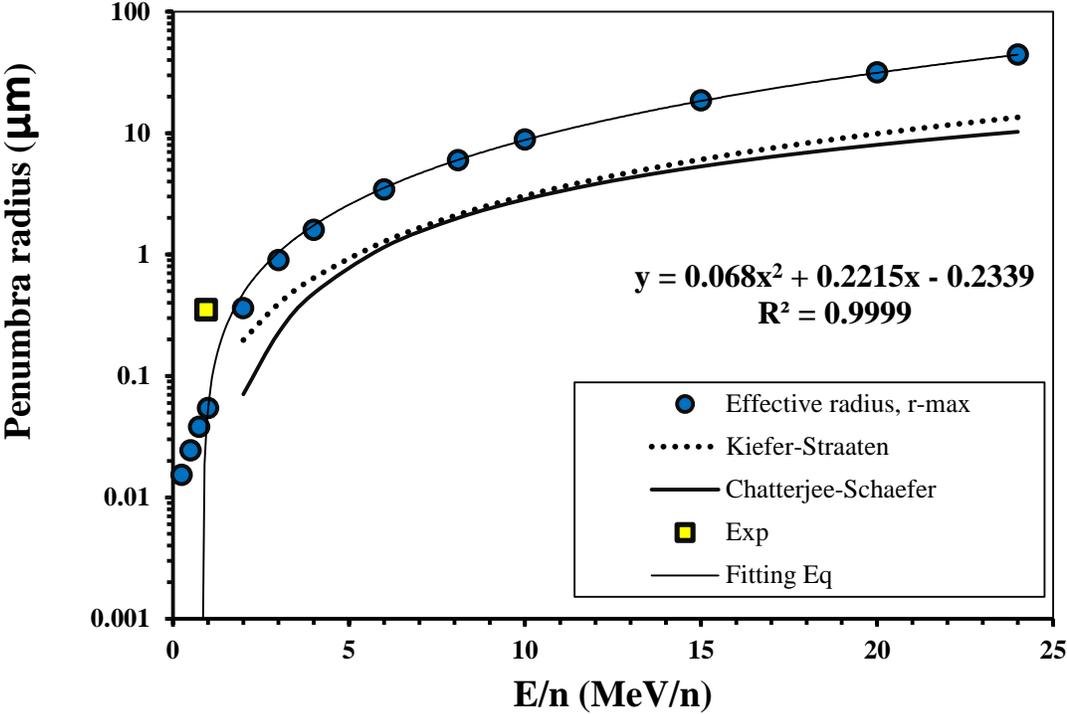

Table 1

| $i$ | $b_i$ | $d_i$ |
|---|---|---|
| 1 | $0.2335 \pm 0.0091$ | $(2.98 \pm 0.3) \times 10^3$ |
| 2 | $1.209 \pm 0.015$ | $6.14 \pm 0.29$ |
| 3 | $(1.78 \pm 0.36) \times 10^{-4}$ | $1.026 \pm 0.02$ |
| 4 | $0.9891 \pm 0.001$ | $(2.57 \pm 0.12) \times 10^2$ |
| 5 | $(3.01 \pm 0.35) \times 10^{-4}$ | $0.34 \pm 0.19$ |
| 6 | $1.468 \pm 0.09$ | $(1.47 \pm 0.19) \times 10^3$ |
| 7 | $(1.18 \pm 0.097) \times 10^{-2}$ | $0.692 \pm 0.039$ |
| 8 | $1.232 \pm 0.067$ | $0.905 \pm 0.031$ |
| 9 | $0.109 \pm 0.017$ | $0.1874 \pm 0.0086$ |

Table 2

| E/n (MeV/n) | β | Studied bands of ions |
|---|---|---|
| 0.25 | 0.02316 | P, $\alpha^{*,**}$, C, O, $I^*$ |
| 0.50 | 0.03275 | P, $D^*$, $\alpha$, C, O, $Br^*$, $I^*$ |
| 0.75 | 0.04010 | P, $\alpha^{*,**}$, C, O, Fe |
| 1.00 | 0.046 | $P^{*,**}$, $D^*$, $\alpha$, C, O, Ne, Fe |
| 2.00 | 0.06542 | $P^*$, $\alpha$, $C^{*,**}$, O, Ne, Fe |
| 2.40 | 0.072 | P, $\alpha$, C, $O^*$, Ne, Fe |
| 2.57 | 0.074 | P, $\alpha$, C, $O^{*,**}$, Ne, Fe |
| 3.00 | 0.080 | $P^*$, $\alpha$, C, O, Ne, Fe |
| 4.00 | 0.092376 | P, $\alpha$, C, O, Ne, Fe |
| 6.00 | 0.112956 | P, $\alpha$, C, O, Ne, Fe |
| 8.10 | 0.131023 | P, $\alpha$, C, O, $Ne^*$, Fe |
| 10.0 | 0.145361 | $P^{**}$, $\alpha^{**}$, $C^{**}$, $O^{**}$, $Fe^{**}$ |
| 15.0 | 0.1773 | P, $\alpha$, O, Fe |
| 20.0 | 0.204 | P, $\alpha$, O, Fe |
| 24.0 | 0.222721 | P, $\alpha$, C, O, Ne, Fe |